\documentclass[10pt,conference]{IEEEtran}
\pdfoutput=1

\usepackage{cite}
\usepackage{amsmath}
\usepackage{graphicx}
\usepackage{textcomp}
\usepackage{booktabs}
\usepackage{multirow}
\usepackage{multicol}

\newcommand{\abs}[1]{\left\lvert#1\right\rvert}

\PassOptionsToPackage{hyphens}{url}\usepackage{hyperref}
\usepackage{scalerel}
\usepackage{tikz}
\usetikzlibrary{svg.path}

\definecolor{orcidlogocol}{HTML}{A6CE39}
\tikzset{
  orcidlogo/.pic={
    \fill[orcidlogocol] svg{M256,128c0,70.7-57.3,128-128,128C57.3,256,0,198.7,0,128C0,57.3,57.3,0,128,0C198.7,0,256,57.3,256,128z};
    \fill[white] svg{M86.3,186.2H70.9V79.1h15.4v48.4V186.2z}
                 svg{M108.9,79.1h41.6c39.6,0,57,28.3,57,53.6c0,27.5-21.5,53.6-56.8,53.6h-41.8V79.1z M124.3,172.4h24.5c34.9,0,42.9-26.5,42.9-39.7c0-21.5-13.7-39.7-43.7-39.7h-23.7V172.4z}
                 svg{M88.7,56.8c0,5.5-4.5,10.1-10.1,10.1c-5.6,0-10.1-4.6-10.1-10.1c0-5.6,4.5-10.1,10.1-10.1C84.2,46.7,88.7,51.3,88.7,56.8z};
  }
}

\newcommand\orcidicon[1]{\href{https://orcid.org/#1}{\mbox{\scalerel*{
\begin{tikzpicture}[yscale=-1,transform shape]
\pic{orcidlogo};
\end{tikzpicture}
}{|}}}}

\begin{document}

\title{Feature Selection for Classification with QAOA\\
}

\author{\IEEEauthorblockN{1\textsuperscript{st} Gloria Turati
\orcidicon{0000-0001-5367-4058}}
\IEEEauthorblockA{
\textit{Politecnico di Milano}\\
Milano, Italy \\
gloria.turati@polimi.it}
\and
\IEEEauthorblockN{2\textsuperscript{nd} Maurizio {Ferrari Dacrema} \orcidicon{0000-0001-7103-2788}}
\IEEEauthorblockA{\textit{Politecnico di Milano}\\
Milano, Italy \\
maurizio.ferrari@polimi.it}
\and
\IEEEauthorblockN{3\textsuperscript{rd} Paolo Cremonesi \orcidicon{0000-0002-1253-8081}}
\IEEEauthorblockA{
\textit{Politecnico di Milano}\\
Milano, Italy \\
paolo.cremonesi@polimi.it}
}

\maketitle

\begin{abstract}
Feature selection is of great importance in Machine Learning, where it can be used to reduce the dimensionality of classification, ranking and prediction problems. The removal of redundant and noisy features can improve both the accuracy and scalability of the trained models. However, feature selection is a computationally expensive task with a solution space that grows combinatorically. In this work, we consider in particular a quadratic feature selection problem that can be tackled with the Quantum Approximate Optimization Algorithm (QAOA), already employed in combinatorial optimization. First we represent the feature selection problem with the QUBO formulation, which is then mapped to an Ising spin Hamiltonian. Then we apply QAOA with the goal of finding the ground state of this Hamiltonian, which corresponds to the optimal selection of features.
In our experiments, we consider seven different real--world datasets with dimensionality up to 21 and run QAOA on both a quantum simulator and, for small datasets, the 7--qubit IBM (ibm--perth) quantum computer.
We use the set of selected features to train a classification model and evaluate its accuracy.
Our analysis shows that it is possible to tackle the feature selection problem with QAOA and that currently available quantum devices can be used effectively. 
Future studies could test a wider range of classification models as well as improve the effectiveness of QAOA by exploring better performing optimizers for its classical step.
\end{abstract}


\section{Poster Relevance}

In this work we address the feature selection problem for classification using the Quantum Approximate Optimization Algorithm (QAOA). We observe that the final classifiers are comparable and sometimes outperform the classifiers obtained by performing feature selection with classical algorithms and Quantum Annealers (QA).

\section{Extended Poster Abstract}

\subsection{Introduction}

In recent years, the spread of big data and Machine Learning has led to the need for methods that facilitate the understanding, analysis and processing of information. This is especially important when data are used for classification, ranking and prediction tasks, where the redundancy of information can lead to computationally expensive algorithms and cause overfitting problems.
Feature selection is useful for this purpose, as the elimination of irrelevant features allows to speed up the training of the model and improve its accuracy.
However, the process of deciding which features are the most relevant is computationally expensive: the direct evaluation of all possible subsets of features is an NP--hard problem and its complexity increases factorially with the number of the features.
The techniques traditionally used to perform feature selection can be summarized in three main categories: filter, wrapper and embedded methods, which have been empirically shown to be effective, especially when combined together, but still have a high computational cost\cite{DBLP:journals/cee/ChandrashekarS14}\cite{DBLP:conf/mipro/JovicBB15}.

Some papers have already tackled feature selection with quantum computing with the goal of addressing its classical limitations.
\cite{DBLP:conf/sigir/DacremaMN0FC22} has shown that QA allows to find features producing classification models with accuracy comparable to classical solvers, by testing three different feature selection approaches on 15 publicly available datasets.
Another paper by \cite{DBLP:journals/corr/abs-2203-13261} has focused on Mutual Information and shown similar results compared to \cite{DBLP:conf/sigir/DacremaMN0FC22}; the analysis includes the Variational Quantum Eigensolver (VQE) but the algorithm is only reported for a single synthetically generated dataset.
In this paper we do a step forward and report preliminary results for QAOA on three feature selection approaches on seven real--world datasets.

\subsection{Problem Formulation}\label{sec2}

We consider a classification problem with $n$ features $f_1,\dots,f_n\in F$ and target variable $y$: the goal is to select the most informative $k<n$ features with which to feed the classification algorithm.
We treat the feature selection task as a Quadratic Unconstrained Binary Optimization (QUBO) \cite{DBLP:journals/anor/GloverKHD22} problem
\begin{equation}
    \quad \min_{x\in\{0,1\}^n} x^T Q  x
\end{equation}
where $x=(x_1,\dots,x_n)\in \{0,1\}^n$ and $Q$ is an upper triangular or symmetrical $n \times n$ matrix.
The binary variables $x$ represent the solution of the feature selection problem ($x_i = 1$ if and only if the $i$--th feature is selected) and $Q$ is defined in such a way that the cost function represents the amount of information provided by the selected features. We consider three QUBO models characterized by different definitions of the objective function.

\begin{subsubsection}{QUBO Correlation}
This approach aims to select features highly correlated with the target label. Here the selection of features correlated to each other is discouraged to avoid redundancy. Thus we define
    \begin{equation}
    Q_{i,j} =
    \begin{cases}
    r(f_i,f_j) \quad & \text{if $i\neq j$} \\
    -r(f_i,y) \quad & \text{if $i=j$}  
    \end{cases}
    \end{equation}
where $r(\cdot,\cdot)$ is the Pearson correlation coefficient.
\end{subsubsection}

\begin{subsubsection}{QUBO Mutual Information}Here the QUBO matrix is built as
    \begin{equation}
    Q_{i,j} =
    \begin{cases}
    -MI(f_i,y|f_j) \quad & \text{if $i\neq j$} \\
    -MI(f_i,y) \quad & \text{if $i=j$}  
    \end{cases}
    \end{equation}
    where $MI(f_i,y)$ is the mutual information between the $i$--th feature and the target variable $y$ and $MI(f_i,y|f_j)$ is the conditional mutual information between $f_i$ and $y$, assuming that $f_j$ is known.
\end{subsubsection}

\begin{subsubsection}{QUBO Boosting}
    This metric makes use of a Support Vector Classifier (SVC) to estimate the level of information provided by a feature. It is built by first considering each of the $\abs{F}$ features and training a support vector classifier (SVC) for each of them; each of the trained classifiers is then used to predict the outcome variable. If we denote by $h_i$ the class predicted using the model trained considering only the $i$--th feature, we can define
    \begin{equation}
    Q_{i,j} =
    \begin{cases}
    r(h_i,h_j) \quad & \text{if $i\neq j$} \\
    \frac{S}{\abs{F}^2} + \lambda - 2 \cdot r(h_i,y) \quad & \text{if $i=j$} 
    \end{cases}
    \end{equation}
    where $\lambda$ is a hyper--parameter, $S$ is the number of samples in the dataset, $r(h_i,h_j)$ is the Pearson correlation coefficient between labels predicted with different models and $r(h_i,y)$ is the correlation between each predicted label and the correct one (see \cite{DBLP:journals/jmlr/NevenDRM12} for further details).
\end{subsubsection}

At this point, in order to control the number of selected features, we add a quadratic penalty term to the cost function so that its value increases when the number of actually selected features deviates from the target $k$. The resulting minimization problem is
\begin{equation}
    \min_{x\in\{0,1\}^n} x^T Q  x + \left(\sum_{i=1}^{n}x_i - k\right)^2
\end{equation}
Once our problem has been written in the QUBO formulation, we can convert it to the Ising formulation \cite{DBLP:journals/corr/abs-1302-5843} and then apply QAOA.

\subsection{Selecting Features with QAOA}
Differently from other Variational Quantum Algorithms, QAOA has a specific circuit, which depends on the parameters $\gamma=(\gamma_1,\dots,\gamma_p)$ and $\beta=(\beta_1,\dots,\beta_p)$. It consists of $p$ layers where gates implementing the cost operator $U_C(\gamma)$ and the mixer operator $U_M(\beta)$ are applied (see the original work by Farhi for further details on the structure of QAOA circuit \cite{https://doi.org/10.48550/arxiv.1411.4028}).
QAOA algorithm involves a number of iterations, in each of which the parameters are fixed and the cost function is estimated by executing the QAOA circuit $m$ times, measuring the final states and averaging the expectations of the Hamiltonian on the collapsed states.
At the end of each iteration, a classical optimizer modifies the parameters with the goal of minimizing the cost function.
The algorithm stops when the parameters meet certain tolerance criteria, specified in the construction of the optimizer (we will not analyze the stopping criterion, as the behaviour of the optimizer is outside the scope of our work).
Once the optimal parameters have been found, we run the final circuit $m$ times again and pick the solution with the lowest cost, which represents the selection of the most relevant features.

\begin{table*}[htbp]
  \centering
  \caption{Accuracy for all QUBO methods, datasets and solvers. The best accuracy for each QUBO method and dataset is highlighted in bold. Experiments with the IBM QPU are performed only for datasets with no more than seven features}
    \begin{tabular}{clr|cc|cc|ccc}
    \toprule
      &       &       & \multicolumn{2}{|c}{QAOA} & \multicolumn{2}{|c}{QA} & \multicolumn{3}{|c}{Traditional Solver} \\
    Method & \multicolumn{1}{c}{Dataset} & \multicolumn{1}{c}{n} & \multicolumn{1}{|c}{Simulator} & \multicolumn{1}{c}{IBM QPU} & \multicolumn{1}{|c}{D--Wave QPU} & \multicolumn{1}{c}{QPU Hybrid} & \multicolumn{1}{|c}{SA} & \multicolumn{1}{c}{SD} & \multicolumn{1}{c}{TS} \\
    \midrule
    \multicolumn{1}{c}{\multirow{7}[0]{*}{\parbox{2cm}{\centering{ QUBO\\ Correlation}}}} & iris  & 4     & 0.9022 & 0.9111 & \textbf{0.9333} & 0.9289 & \textbf{0.9333} & \textbf{0.9333} & 0.9289 \\
          & cars1 & 7     & 0.8424 & 0.8458 & 0.8373 & 0.8458 & 0.8458 & \textbf{0.8508} & 0.8458 \\
          & LED--display--domain--7digit & 7     & 0.5467 & 0.5467 & 0.5333 & 0.5413 & 0.5427 & 0.5440 & \textbf{0.6613} \\
          & breast--cancer & 9     & 0.6674 & -     & 0.6884 & 0.6791 & 0.6953 & 0.6698 & \textbf{0.7070} \\
          & wine  & 13    & 0.9407 & -     & 0.8852 & 0.8815 & 0.8778 & 0.8852 & \textbf{0.9481} \\
          & vehicle & 18    & 0.7339 & -     & 0.7118 & 0.7197 & 0.7063 & 0.7102 & \textbf{0.7425} \\
          & thyroid--ann & 21    & \textbf{0.9898} & -     & 0.9470 & 0.9473 & 0.9468 & 0.9475 & 0.9484 \\
    \midrule
    \multicolumn{1}{c}{\multirow{7}[0]{*}{\parbox{2cm}{\centering{ QUBO\\ Mutual \\ Information}}}} & iris  & 4     & 0.8756 & 0.8844 & 0.9111 & 0.9111 & 0.9111 & 0.9111 & \textbf{0.9333} \\
          & cars1 & 7     & 0.8441 & 0.8407 & 0.8424 & 0.8424 & \textbf{0.8492} & 0.8441 & 0.8390 \\
          & LED--display--domain--7digit & 7     & \textbf{0.6787} & 0.6747 & \textbf{0.6787} & 0.6760 & \textbf{0.6787} & 0.6720 & 0.6773 \\
          & breast--cancer & 9     & \textbf{0.7302} & -     & 0.6791 & 0.6907 & 0.6860 & 0.6860 & 0.7233 \\
          & wine  & 13    & 0.9630 & -     & 0.9556 & 0.9630 & 0.9630 & 0.9667 & \textbf{0.9741} \\
          & vehicle & 18    & 0.7591 & -     & 0.7283 & 0.7409 & 0.7386 & 0.7402 & \textbf{0.7606} \\
          & thyroid--ann & 21    & 0.9890 & -     & 0.9249 & 0.9249 & 0.9249 & 0.9249 & \textbf{0.9929} \\
    \midrule
    \multicolumn{1}{c}{\multirow{7}[0]{*}{\parbox{2cm}{\centering{ QUBO\\ SVC\\ Boosting}}}} & iris  & 4     & \textbf{0.9333} & \textbf{0.9333} & \textbf{0.9333} & \textbf{0.9333} & \textbf{0.9333} & 0.9244 & \textbf{0.9333} \\
          & cars1 & 7     & \textbf{0.8271} & 0.7695 & 0.7203 & 0.7220 & 0.7220 & 0.7237 & 0.7254 \\
          & LED--display--domain--7digit & 7     & \textbf{0.5480} & 0.4373 & 0.4000 & 0.4000 & 0.4000 & 0.4000 & 0.3947 \\
          & breast--cancer & 9     & \textbf{0.7070} & -     & \textbf{0.7070} & 0.6860 & 0.6465 & 0.6651 & 0.6488 \\
          & wine  & 13    & 0.8778 & -     & 0.9370 & 0.9370 & 0.9333 & 0.9370 & \textbf{0.9556} \\
          & vehicle & 18    & 0.7134 & -     & \textbf{0.7386} & 0.6866 & 0.6953 & 0.6913 & 0.7307 \\
          & thyroid--ann & 21    & \textbf{0.9811} & -     & 0.9661 & 0.9703 & 0.9666 & 0.9693 & 0.9684 \\
    \bottomrule
    \end{tabular}%
    \label{table1}
\end{table*}%

\subsection{Experiments and Results}

We perform two experiments with the goal of:
\begin{itemize}
    \item comparing QAOA with Quantum Annealers (QA) and classical solvers when used to select features for training classification models
    \item studying the dependence of the final classification accuracy on the depth $p$ of the circuit.
\end{itemize}

\subsubsection{Comparison of QAOA with QA and classical solvers}

We consider the seven real--world datasets ``iris'', ``cars1'', ``LED--display--domain--7digit'', ``breast--cancer'', ``wine'', ``vehicle'', ``thyroid--ann'', taken from OpenML \cite{DBLP:journals/sigkdd/VanschorenRBT13}, and having $n=4,7,7,9,13,18,21$ features respectively. Then we define the $Q$ matrix according to the QUBO formulations described in \ref{sec2} and choose $k=\lceil 50\% \, n \rceil$ as the target number of features to be selected.
We run QAOA for all the possible combinations of datasets and methods using the IBM ``QASM simulator'' and consider circuits having depth $p$ ranging from 1 to 10.
The number of shots used for approximating the cost function and for the final sampling is fixed to 100.
Finally, we choose the SciPy ``COBYLA'' as a classical optimizer, which is gradient--free and has been shown to have good performances when the environment noise is negligible \cite{DBLP:conf/qce/LavrijsenTMIJ20}.
In addition, for the three smallest datasets, we run QAOA using also the real 7--qubit ``ibm--perth'' QPU.
We still perform 100 shots for each iteration and choose the ``COBYLA'' optimizer, but this time we only consider the circuit depths $p$ ranging from 1 to 6, due to the long queue times to access the QPU.

The optimization of the QUBO functions and the selection of features with Quantum Annealers and classical solvers are performed as described in \cite{DBLP:conf/sigir/DacremaMN0FC22}, using the publicly available code. In particular, experiments on Annealers are conducted on the D--Wave 5600--qubit Advantage QPU and with a hybrid quantum--classical approach consisting in decomposing the QUBO problem in smaller tasks which are then executed on the Annealer.
The traditional solvers tested are Simulated Annealing (SA), Steepest Descent (SD) and Tabu Search (TS).

At this point, we use the selected features to train the classifier. We choose Random Forest as a classification model, since it has been shown to provide good accuracy even with limited tuning.
The datasets used are split into training and test set with $70\%$ and $30\%$ of the samples respectively.
Since some datasets have limited size, we do not create an additional subdivision for validation, but apply a 5--Fold CV on the training set and average the scores.
For solvers making use of QAOA, the hyper--parameter $p$ with the best score on the validation is chosen.
Finally, we evaluate the model on the test set.
Since Random Forest has an inherent stochasticity, we perform the evaluation 5 times and average the accuracy obtained.
The results are summarized in Table~\ref{table1}.

\begin{table*}[h!]
\centering
   \caption{Average accuracy and standard deviations of the classification models for the dataset ``thyroid--ann'' obtained with the features selected with all QUBO methods and $p=1,\dots,10$. The error bars in the plot represent the standard deviation}
    	\begin{minipage}{0.5\linewidth}
    	\centering
            \begin{tabular}{rrrr}
            \toprule
                p     & \parbox{1.9cm}{\centering{ QUBO\\ Correlation}} & \parbox{1.9cm}{\centering{ QUBO\\ Mutual\\ Information}} & \parbox{1.9cm}{\centering{ QUBO\\ SVC\\ Boosting}} \\
                \midrule
                1     & 0.9766 ± 0.0093 & 0.9657 ± 0.0192 & 0.9722 ± 0.0076 \\
                2     & 0.9717 ± 0.0090 & 0.9323 ± 0.0075 & 0.9637 ± 0.0106 \\
                3     & 0.9422 ± 0.0236 & 0.9666 ± 0.0253 & 0.9623 ± 0.0084 \\
                4     & 0.9701 ± 0.0183 & 0.9567 ± 0.0261 & 0.9669 ± 0.0113 \\
                5     & 0.9639 ± 0.0262 & 0.9583 ± 0.0281 & 0.9609 ± 0.0138 \\
                6     & 0.9575 ± 0.0278 & 0.9620 ± 0.0200 & 0.9604 ± 0.0096 \\
                7     & 0.9468 ± 0.0280 & 0.9505 ± 0.0276 & 0.9685 ± 0.0120 \\
                8     & 0.9765 ± 0.0088 & 0.9653 ± 0.0296 & 0.9683 ± 0.0061 \\
                9     & 0.9535 ± 0.0304 & 0.9606 ± 0.0247 & 0.9583 ± 0.0265 \\
                10    & 0.9441 ± 0.0302 & 0.9639 ± 0.0250 & 0.9701 ± 0.0152 \\
            \bottomrule
            \end{tabular}%
        \end{minipage}\hfill
    	
    	\begin{minipage}{0.45\linewidth}
    		\centerline{\includegraphics[width=12cm]{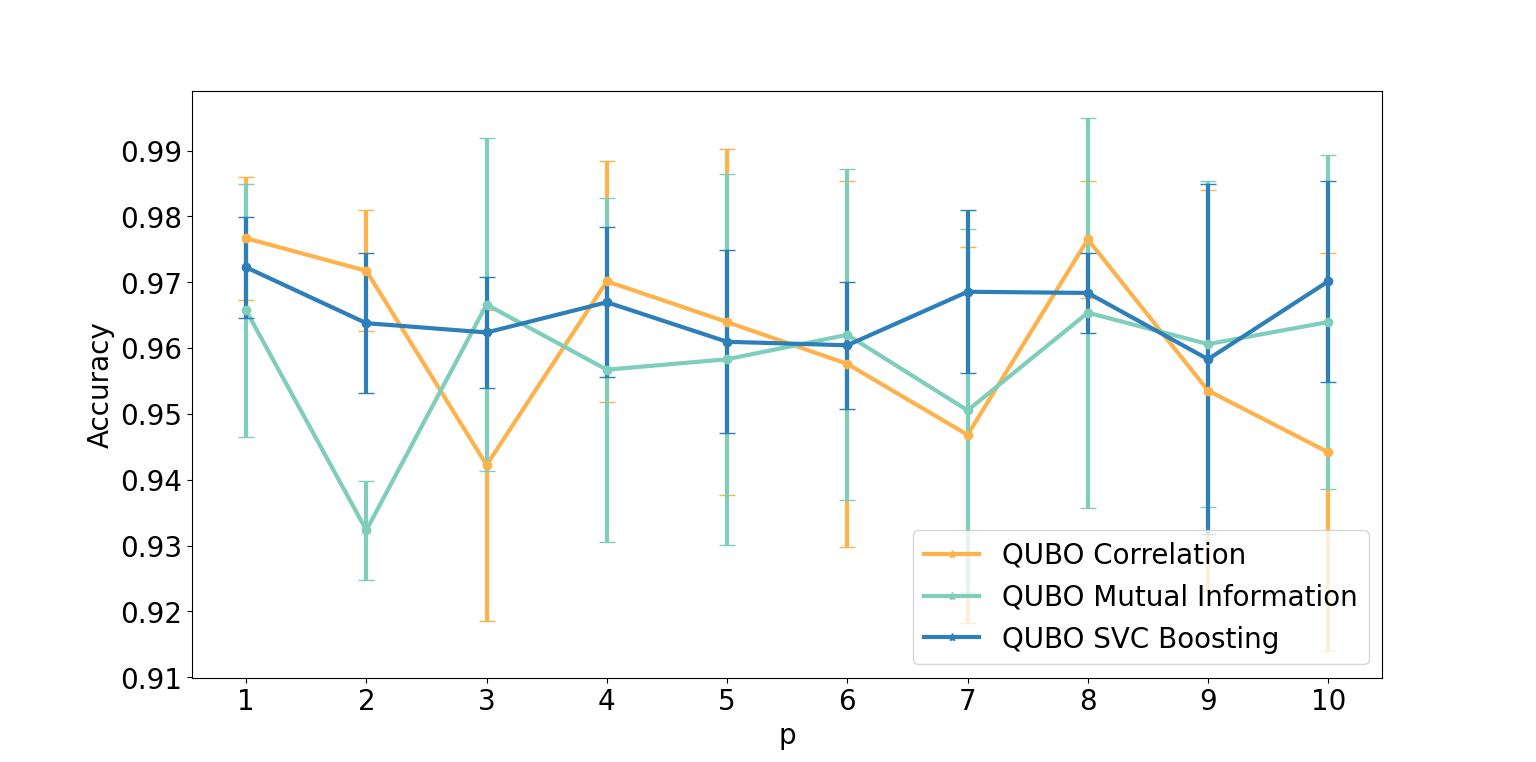}
            }
    	\end{minipage}
    	\label{table2}
\end{table*}

It can be observed that the accuracy of the model trained using features selected with QAOA is comparable to that obtained by selecting features with Quantum Annealers and traditional solvers.
In particular, there is no QUBO formulation that outperforms the others and the accuracy is strictly dependent on the dataset.
Finally, the results obtained by running QAOA on the simulator are very similar to those provided by the real QPU, an indication that also the quantum hardware is reliable in solving the feature selection problem for classification.

\subsubsection{Dependence of the accuracy on $p$}
In order to study the dependence of the accuracy on the depth of the circuit $p$, we consider the same datasets and QUBO formulations as the previous experiment and run QAOA on the IBM ``QASM simulator'' for all $p$ in the range from 1 to 10. We still perform 100 executions of the circuit per iteration and choose ``COBYLA'' as a classical optimizer. Once we get the most relevant features, we train the Random Forest and compute the accuracy. We execute this pipeline 5 times and average the accuracy.

Due to space reasons, we report in Table~\ref{table2} only the results for the dataset ``thyroid--ann'', which is the largest we tested. The behaviour of $p$ in the other datasets is similar.
The table and the corresponding graph show the average accuracy and the standard deviation of the final classifier.
It can be observed that the accuracy varies with $p$, but, due to the high standard deviation (with a mean of 0.0191), we can only conclude that $p$ is a challenging hyper--parameter to tune.

\subsection{Conclusions and Future Work}

This work shows that it is possible to perform feature selection with QAOA and obtain classification accuracy comparable to what obtained by selecting features with classical approaches.
However, further studies may be conducted to test a wider range of classification models and analyze the behaviour of the QPU on datasets with a larger number of features.
Furthermore, the effectiveness of QAOA may be improved by exploring other optimizers which may be better suited for specific feature selection heuristics or to the noisy nature of the QPU.

\bibliographystyle{IEEEtran}
\bibliography{IEEEabrv,QCE-bibliography}

\end{document}